\documentstyle[12pt,epsfig]{article}

\def\Fe{{\rm Fe}}
\def\Ni{{\rm Ni}}
\def\Co{{\rm Co}}

\def\roughly#1{\mathrel{\raise.3ex\hbox{$#1$\kern-.75em%
\lower1ex\hbox{$\sim$}}}}
\def\fm{{\rm fm}}
\def\lsim{\roughly<}

\def\be{\begin{eqnarray}}
\def\ee{\end{eqnarray}}
\def\Tr{{\rm Tr}\;}

\def\ben{\begin{enumerate}}
\def\een{\end{enumerate}}
\def\beitem{\begin{itemize}}
\def\eitem{\end{itemize}}
\def\fm{{\rm fm}}

\newcommand{\beq}{\begin{eqnarray}}
\newcommand{\eeq}{\end{eqnarray}}

\def\bi{\begin{itemize}}
\def\ei{\end{itemize}}

\def\etal{{\it et al}}
\def\del{\partial}
\def\L{{\cal L}}

\long\def\beginomit#1\endomit{}
\def\np{{ Nucl. Phys.}}
\def\prl{{ Phys. Rev. Lett.}}

\def\pl{{ Phys. Lett.}}
\def\L{{\cal L}}

\def\AJ{{ Astrophys. J.}}
\def\AA{{ Astron. \& Astrophys.}}
\def\PR{{ Phys. Repts.}}
\def\chpt{$\chi$PT}
\def\be{\begin{eqnarray}}
\def\ee{\end{eqnarray}}
\def\Tr{{\rm Tr}\;}

\def\fm{{\rm fm}}
\def\etal{{\it et al.}}

\def\del{\partial}
\def\L{{\cal L}}
\def\M{{\cal M}}
\def\MeV{{\mbox MeV}}
\def\M{{\cal M}}

\begin{document}

%-------------------------------------------------------------------------%
%                     Title Page                                          %
%-------------------------------------------------------------------------%

\begin{center}

\hfill{SNUTP-95-036}

\hfill{hep-ph/9503382}

\hfill{March 1995}
\vskip 0.4in

{\Large\bf Kaon Condensation in Dense Stellar Matter
\footnote{Based on
talks given by CHL and MR at International Workshop on Nuclear and
Particle Physics, {\it ``Chiral Dynamics in Hadrons and Nuclei"},
Feb. 6 $\sim$ Feb.10, 1995, Seoul Nat'l University, Seoul, Korea}
}
\vskip 1cm
{\bf CHANG-HWAN LEE}\\
{ \it Center for Theoretical Physics and Department of Physics } \\
{ \it Seoul National University, Seoul 151-742, Korea}\\
\vskip .4cm

and

\vskip .4cm
{\bf  MANNQUE RHO}\\
{ \it Service de Physique Th\'{e}orique, CEA  Saclay}\\
{\it 91191 Gif-sur-Yvette Cedex, France}\\
{and}\\
{\it Institute for Nuclear Theory, University of Washington}\\
{\it Seattle, WA 98195, U.S.A.}
\vskip 1.5cm

\begin{abstract}
This article combines two talks given by the authors and is based on
works done in collaboration with
G.E. Brown and D.-P. Min on kaon condensation in dense baryonic medium treated
in chiral perturbation theory using heavy-baryon formalism.
It  contains,
in addition to what was recently published,
astrophysical backgrounds for kaon condensation discussed by Brown and Bethe,
a discussion on a renormalization-group analysis
to meson condensation worked out together with H.K. Lee and S.-J. Sin,
and the recent results of K. M. Westerberg in the bound-state approach to the
Skyrme model. Negatively charged
kaons are predicted to condense at a
critical density $2\lsim \rho/\rho_0\lsim 4$, in the range to allow the
intriguing new phenomena predicted by Brown and Bethe to take place
in compact star matter.
\end{abstract}
\end{center}

\section{Motivation}

Recent work by Bethe and Brown\cite{BB} on the maximum mass of stable
compact stars --
called ``neutron stars" in the past but more appropriately
``nuclear (or nucleon) stars" -- suggest that the nuclear equation of state
(EOS) in the interior of compact stars must be considerably softened
at densities a few times the nuclear matter density $\rho_0$ by one or
several hadronic phase transitions. It is now fairly clear that
neither pion condensation nor quark matter will figure at a density low enough
to be relevant to the star matter although the issue is not yet
completely settled.
As Bethe and Brown suggest, kaon condensation could however take place
at a density 3--4 times the normal matter density and hence play an
important role in explaining the remarkably
narrow range of compact star masses observed in nature\cite{mass}.

The aim of this talk is to describe a higher-order chiral perturbation
calculation that predicts
the critical density for kaon condensation. The strategy is to take up
what Kaplan and Nelson\cite{KN} started, namely chiral perturbation
theory (\chpt). \ Kaplan and Nelson predicted in tree order of \chpt \ that
kaons condense in neutron matter at $\rho\lsim 3\rho_0$. Our calculation
goes to next-to-next-to-leading (NNL) order. It turns out that
the calculation confirms the Kaplan-Nelson prediction although in the
process new and interesting physical elements are uncovered. Our result is that
for reasonable ranges of parameters involved, the critical density
comes out to be
\be
2\lsim \rho/\rho_0 \lsim 4.
\ee
This is the range of densities relevant to the Bethe-Brown scenario for the
formation of light-mass black holes for stars that exceed the
critical mass of $M\simeq 1.56 M_\odot$. Their arguments extend the estimated
range of main sequence star masses, for which stars go into black holes,
down to $\sim 18 M_\odot$.

There are two situations where the production of kaons brings out
interesting physics. One is their properties in relativistic heavy-ion
collisions that involve temperature. Here kaon condensation is
not directly relevant but the {\it mechanism} that triggers kaon condensation
in the relevant situation has intriguing consequences on the properties
of kaons observed in heavy-ion experiments. This is discussed in a recent
review\cite{newbr94} and will not be discussed here. What we are interested
in is what kaons do in cold dense matter appropriate to compact objects
that result from the collapse of massive stars.

In stellar collapse, as matter density $\rho$ increases, the electron
chemical potential $\mu_e$ (determined by the chemical potentials of neutrons
and protons in the system together with charge neutrality)
increases, reaching several hundreds of MeV.
If the electron chemical potential reaches the ``effective mass" of
a meson $\Phi$, $m_\Phi$, then the electron can ``decay" into a $\Phi$
as\cite{BKR}
\be
e^-\rightarrow \Phi^- +\nu_e.\label{edecay}
\ee
In nature, the only low-mass bosons are the pseudo-Goldstone bosons
$\Phi^-=\pi^-, K^-$. While lowest in mass, the pions do not seem to play
an important role, so the next possible boson is the kaon with
its mass $\sim 500$ MeV in free space. The electron chemical potential
cannot reach this high, so on-shell kaons cannot be produced by this process.
However as will be described below, the kaon in medium can undergo a mass
shift due to density-dependent renormalization. As the $\mu_e$ increases
and the effective kaon mass
$M_K^\star$ decreases as $\rho$ increases, the process (\ref{edecay})
can occur at some density $\rho_c$. Kaons so produced will bose-condense
at that density $\rho_c$. Whether or not this will occur then depends on
whether or not $M_K^\star$ will decrease enough in density so that
it meets $\mu_e$. Such a condensation will be of physical interest if
the critical density is low enough and the energy gain is high enough.
This is the possibility we shall address below.
%--------------------------------------------------------------

\section{Maximum Neutron Star Mass}
%-----------------------------------------------------------------------
\subsection{\it Stellar Death Function}
\begin{figure}
\centerline{\epsfig{file=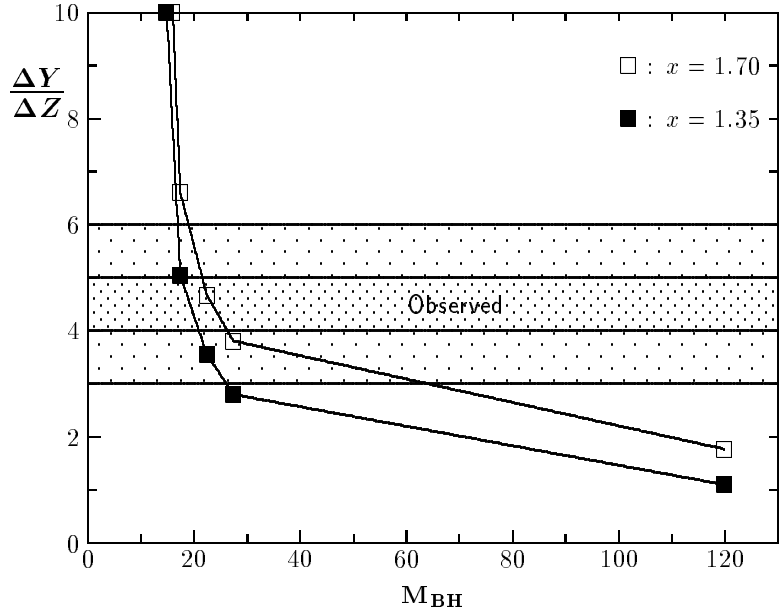,height=8cm}}
\caption[death]{Values of the ratio $\Delta Y/\Delta Z$ of the relative
helium to metal enrichments for different values of x as defined
in eq.(\ref{maeder}) as a function of ${\rm M}_{\rm BH}$.
The data points correspond to initial metallicity Z=0.001.  The
observed range is indicated by shadings; the range 4 to 5 is the
preferred one.\protect{\cite{Maeder}} }
\label{deathfunc}
\end{figure}

That main sequence stars of mass  $\ge 25 -30 M_\odot$ must end up in black
holes without producing nucleosynthesis, i.e., without returning matter to
the galaxy, is required by the observed abundances of
elements\cite{GEB94,Maeder}.
Maeder's argument is based on the measurement of $\Delta Y/\Delta Z$, the
ratio of helium abundance to that of metals, in low-metallicity extragalactic
$H_{II}$ regions, especially irregular dwarf galaxies.
The ratio can be measured with good accuracy\cite{GEB94},
\be
\frac{\Delta Y}{\Delta Z} = 4\pm 1.3 .
\ee
If all stable stars of mass up to $\sim 100 M_\odot$ were to explode, returning
matter to the galaxy, this ratio would lie between 1 and 2. Helium is produced
chiefly by relatively light stars, metals by heavy stars, so that cutting off
the production by the heavy stars going directly into black holes without
nucleosynthesis increases the $\Delta Y/\Delta Z$.
Using the standard initial mass function for stars,
\be
dN/dM=M^{-(1+x)}\label{maeder}
\ee
with $x=1.35, 1.70$,
Maeder\cite{Maeder} found that Pagel's\cite{Pagel}
measurement on $\Delta Y/\Delta Z$ was best reproduced
by a cutoff of nucleosynthesis
at a main sequence stellar mass of $\sim 22.5 M_\odot$ as shown in
Fig.~\ref{deathfunc}.

There is considerable
uncertainty in the initial mass function, as noted by Maeder\cite{Maeder},
so that this limit could easily be $\sim 30 M_\odot$ or even higher.
Brown and Bethe estimated \cite{BB} $30M_\odot$ as the cutoff
for stars to drop directly into black holes without nucleosynthesis.

What about stars with mass $20 M_\odot < M <30 M_\odot$ ?
Recent observation on SN1987A, whose progenitor mass is $\sim 18\pm 2 M_\odot$,
gives us an insight.
Based on the empirical analysis, Brown and Bethe also argued that a large
range of stars below this mass, down
to $\sim 18 M_\odot$, can first accomplish nucleosynthesis and then
collapse into black holes.

%-----------------------------------------------------------------------
\subsection{\it SN1987A: Neutron Star or Black Hole?}

SN1987A(February 23, 1987) in the Large Magellanic Cloud is the
nearest and brightest supernova to be observed since SN1604AD (Kepler), and
certainly the most important supernova since SN1054AD, the progenitor of
Crab Nebula. Because of its brightness and proximity, it will be
possible to observe SN1987A for many years  as it expands to reveal its
inner secrets. In contrast, typical supernova, of which some 20-30
are observed each year, are some 1000 times further and $10^5$ times
fainter, and so become lost in their host galaxy within a year or two.
Moreover,
SN1987A has been observed at every wavelength band of the electromagnetic
spectrum, from radio to $\gamma$-rays.

\begin{figure}
\centerline{\epsfig{file=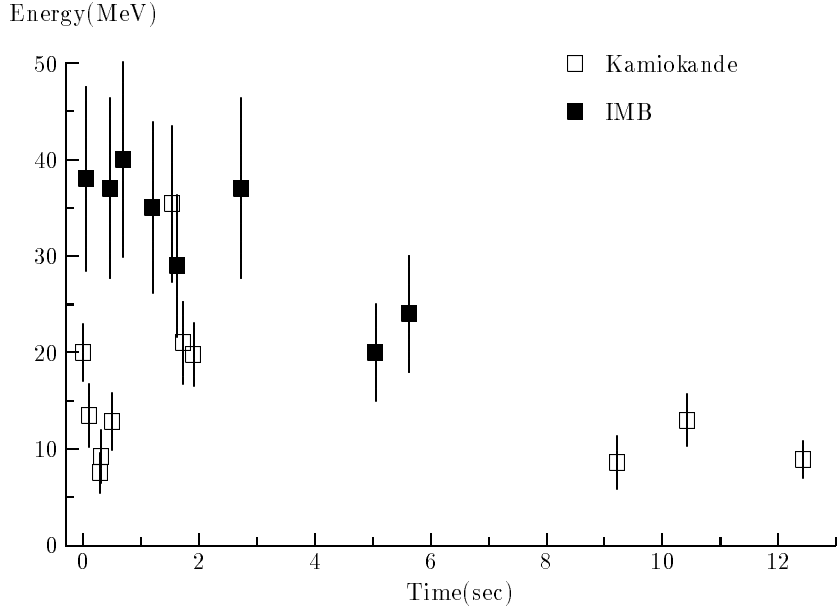,height=8cm}}
\caption[neutrino]{The energy of the neutrinos detected by the Kamiokande
 and IMB detectors is plotted as a single neutrino pulse. Most neutrinos
arrived within the first second or two but there was a lower energy tail to the
pulse which lasted more than 10 sec.}
\label{neutrino}
\end{figure}

In supernova theory, the neutron star is followed by bursts of huge amount
of neutrinos. In the case of SN1987A, neutrinos
were detected by the IMB and Kamiokande detector
about three hours before the optical burst.
The total energy emitted in $\bar\nu_e$'s is
$E_\nu \approx 3\times 10^{52} erg/s$, while
the decay time scale of burst is
$\Delta t\approx 10 \sec$.
These are quite consistent with the expected values.
Hence, firstly, it is believed that the neutron star was formed in SN1987A,
even though there exist time gap of about 7 seconds between the
eighth and ninth Kamiokande events, the gap being followed by another
three events as shown in Fig.~\ref{neutrino}\cite{ENF,SASE}.

{}From the observation of radioactive decay of $^{56}Ni$ and $^{56}Co$,
\be
^{56}\Ni_{28} (\tau_{1/2}=6\, {\mbox{days}})
   &\longrightarrow& ^{56}\Co_{27}+ e^+ +\gamma +\nu_e
\nonumber\\
^{56}\Co_{27}  (\tau_{1/2}=77\, {\mbox{days}})
   &\longrightarrow& ^{56}\Fe_{26}+e^+ + \gamma +\nu_e,
\ee
the mass of Ni ejecta of SN1987A is known to be $0.075 M_\odot$.
Combining this result with the observed energy of SN1987A,
$ E=1.4\pm 0.4 foe $ (where $foe$ stands for $10^{51} erg$),
Brown and Bethe\cite{BB}
obtain the range of the core mass of SN1987A
\be
M =1.535\pm 0.02 M_\odot.
\ee
However, from the mass of the Hulse-Tayler pulsar,
the lower limit of the compact core is known to be at least $1.44 M_\odot$.
Consequently, the compact core mass of SN1987A must range
\be
1.44 M_\odot < M_{core} < 1.56 M_\odot.
\ee

Astronomers have been searching for a
pulsar in the center of SN1987A remnants after the explosion.
To see the explicit signal of a pulsar, one must wait until the remnants
are transparent.
If the pulsar is really formed in SN1987A, the X-ray signal
should be detected within a few years after the explosion.
The hypercritical accretion could hide the compact object for only
$\sim$ 1 year, but after this time, the 1987A should be observed with a
luminosity $L=4\times 10^{38} ergs/s$, but the present light curve
is lower by two orders, $L\sim 4\times 10^{36} ergs/s$\cite{BW94}.
This compact object, after being a proto-neutron star for at least 10 seconds,
appears to have collapsed into a low-mass black hole.
The core of SN1987A may have become a neutron star followed
by neutrino emission,
and later changed into low-mass black hole with mass about $1.5 M_\odot$.

A possible scenario was proposed by Bethe. According to
his arguments, a vigorous convection is produced in the supernova shock.
But after about 2 seconds, the
convection stops as heat is no longer supplied
by neutrinos. As a result, a substantial fraction of the previously
convecting material falls into the neutron star at the center. Bethe estimated
this fraction to be about $10\%$ of the mass in the shock wave, or about
$0.04 M_\odot$. If the neutron-star mass were close to $M_{max}$, the added
$0.04 M_\odot$ could push the neutron star over the limit and make it
collapse into a black hole.
If this happened in SN1987A, it must have been more than 12 seconds after
the first collapse, since at 12 sec, a neutrino was still observed
at Kamiokande II. A
softened equation of state might be associated with
a delayed collapse of the young neutron star into a black hole.

{}From standard evolutionary analysis, it is believed that the neutron star
is formed if the core mass lies
between the maximum neutron star mass($M_{max}$) and
the Chandrasekhar mass,\footnote{
In the early 1930's, Chandrasekhar
set a limit to the size of white dwarf. No carbon-rich white dwarf can
support its weight if it is greater than about 1.4 times solar mass $M_\odot$.
A massive star with its final mass after exhausting its fuel is greater
than $M_{CH}$ collapses and the collapse turns into an explosion.}
\be
M_{CH} = 5.76 Y_e^2 M_\odot,
\ee
where $Y_e$ is the electron fraction per baryon.
In stellar collapse, $Y_e$ is $0.43 \sim 0.50$.
Because of the thermal pressure, the evolutionary lower bound of the
neutron star
is about $(1.10\sim 1.15) \times M_{CH} \approx (1.2\sim 1.4) M_\odot$.

According to Brown and Bethe\cite{BB}, in the binary pulsar evolution,
 the accretion can proceed at the hypercritical rate
\be
\dot{M} \ge 10^4 \dot M_{Edd},
\ee
where the Eddington limit is
\be
\dot M_{Edd} =1.5 \times 10^{-8} M_\odot yr^{-1}.
\ee
Hence, if the neutron star mass were determined by the evolutionary scenario,
massive neutron stars with masses exceeding $1.5 M_\odot$ should exist.  But
as shown below, none have been found.

%-----------------------------------------------------------------------
\subsection{\it Observed Neutron Star Mass}
{} From radio, optical, x-ray and $\gamma$-ray surveys of supernova
remnants, Helfand and Becker came to the conclusion
that nearly half of the supernova in the galaxy leave no observable
remnants. This is understandable because half of the supernova
are Type I, which leave no neutron stars.
However, as discussed in detail by Van den Bergh et al.\cite{BME87},
the observed samples of supernova in our galaxy are at low galactic
latitude, so their light is strongly absorbed on its way to Earth.
They argued that Type II supernova outnumber Type I supernova
by a factor of several. Thus the conclusions of Helfand and Becker
can be understood only if nearly half of the Type II supernova
explosions do not form neutron stars. The possible candidate is, therefore,
a low-mass black hole.

\begin{figure}
\centerline{\epsfig{file=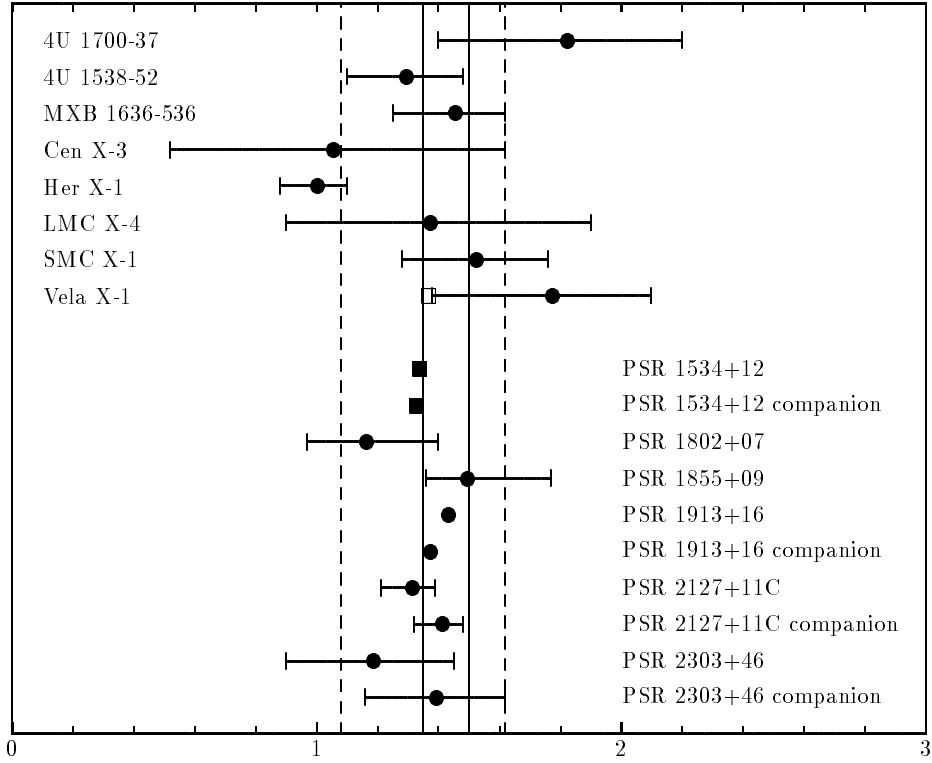,height=10cm}}
\caption[nstar]{Neutron star masses in units of solar mass ($M_\odot$). The
empty box at the lower end
of Vela X-1 is given by the recent analysis of Van
Kerkwijk.\protect{\cite{BB}}}
\label{fignm}
\end{figure}

In Fig.~\ref{fignm}, the measured neutron star masses are plotted.
Most of all the neutron star masses are below $1.5 M_\odot$ except
for Vela X-1 and 4U 1700-37. But by the recent analysis\cite{BB},
the mass of Vela X-1 is believed to be below $1.5 M_\odot$. Further,
there is an argument that 4U 1700-37 is a low-mass black hole\cite{BB}.
If this is the case, it is striking  that all well measured neutron star masses
lie below $1.5 M_\odot$. This calls for theoretical
arguments to lower the maximum neutron star mass. In Table \ref{mmmm}, the
maximum allowed mass of a neutron star is given for various
equations of state\cite{nsmeos}.

\begin{table}
\begin{center}
\begin{tabular}{cc}
\hline
\phantom{aaa} Equation of state \phantom{aaa}
& \phantom{aaa} Maximum mass ($M_\odot$) \phantom{aaa} \\
\hline
 $\pi$ or K condensate & $<$ 1.5 \\
 R & 1.6 \\
BJ & 1.9 \\
TNI & 2.0 \\
TI & 2.0 \\
MF & 2.7 \\
\hline
\end{tabular}
\caption[mm]{The maximum mass of neutron star for various equations of
state cited in \protect{\cite{nsmeos}}}
\label{mmmm}
\end{center}
\end{table}

\subsection{\it Lowering Maximum Neutron Star Mass}
Since the pioneering work by Bahcall and Wolf\cite{bahcall},
it has been known that a pion condensate, if it exists, will enhance
strongly the emissivity of the neutrinos from within the core of a neutron
star. These authors considered the reaction
\be
n+\pi^- \rightarrow n +e^- +\nu
\ee
and its inverse reaction, in which the pions were treated as real particles
existing with a certain probability inside the neutron star. Subsequently
Maxwell et al\cite{MBCDM} carried out a detailed calculation on the
neutrino emissivity in the presence of pion condensates. The reaction
mechanism may be written symbolically as
\be
n+``\pi" \rightarrow n +e^- +\nu
\ee
where $``\pi"$ represents the pion condensate built in the neutron
quasiparticle
states. The main conclusion of \cite{MBCDM} was that even
a small amount of pion condensates will cause a dramatic enhancement
of neutrino emissivity making the equation of state
softer than what standard calculations would predict. This soft equation of
state could then lower the maximum neutron star mass.

However, the much-discussed
P-wave pion condensation is now considered to be rather
unlikely to take place at a low
enough density. Based on renormalization-group flow equations,
Lee et al\cite{LRS}
showed that the Yukawa coupling term responsible for P-wave condensation,
$\bar\psi \vec\tau\cdot \vec\pi \gamma_5 \psi$,
becomes irrelevant after radiative corrections and cannot induce instability
needed for a phase transition.
Furthermore, the axial-vector coupling constant $g_A$ in nuclear
medium is effectively quenched, roughly, down to
one.\cite{rho74}
The quenched axial-vector coupling would then
push the critical density to a higher
density ($>5 \rho_0$),
\be
\rho_c \propto \frac{1}{g_A^2-1},
\ee
where most of the approximations associated with
effective hadronic Lagrangians must have broken down.

An S-wave pion condensation is also unlikely to occur since
chiral symmetry protects the pion mass (PCAC). Even if
S-wave condensation occurred,
the effect would be negligible.

The next candidate process is kaon ($K^-$) condensation.
(The kaon mass in free space is $\sim 495 MeV$).
According to Kaplan and Nelson\cite{KN}, the attraction needed for kaon
condensation comes mainly from the KN sigma term
\be
\delta M_K^2 \approx -\frac{\Sigma_{KN}}{f^2}
\langle N^\dagger N\rangle +\cdots
\ee
where $\Sigma_{KN}$ comes from the explicit chiral symmetry breaking,
\be
\Sigma_{KN} \simeq \frac 12 (\hat m +m_s) \langle N |\bar u u +\bar s s|
N\rangle.
\ee
Here $\hat m$ and $m_s$ are $\sim 5 MeV$ and $\sim 150 MeV$, respectively.
The strangeness content of the proton is not well-determined, so that gives
some uncertainty in the value of $\Sigma_{KN}$. The predicted critical
density was found to be in the range
\be
2.3 \rho_0 \le \rho_c \le 3 \rho_0.
\ee

Recently, an improved calculation was made by
Thorsson et al.\cite{BKRT,TPL}. However this calculation was also incomplete
since the tree-order Lagrangian they used described correctly neither
KN scattering nor kaonic atoms.
In these talks, we describe how this defect is removed in
chiral perturbation theory by going to one-loop order\cite{LJMR,LBM,LBMR,LRp}.
%--------------------------------------------------------------

\section{Baryon Chiral Perturbation Theory}
\subsection{\it Effective Field Theory for Nuclear Matter}
The process we are interested in requires a field theory that can
describe simultaneously normal nuclear matter and phase transitions therefrom.
The most relevant ingredient of QCD that is needed here is
spontaneously broken chiral symmetry. We are specifically interested
in chiral $SU(3)\times SU(3)$ symmetry since strangeness is involved.
In order to address the problem, we need to start from a realistic
effective chiral Lagrangian, obtain a nuclear matter of the right
properties from it and then determine whether strangeness condensation
occurs.

Unfortunately we do not yet know how to describe nuclear matter starting from
a chiral Lagrangian. There are various suggestions and one promising one is
that nuclear matter arises as a solitonic matter from a chiral effective
action, a sort of chiral liquid\cite{lynn} resembling Landau Fermi liquid.
The hope is that the resulting effective action would look like
Walecka's mean-field model. There is as yet no convincing derivation along
this line. However as argued in \cite{newbr94}, there is a compelling
phenomenological indication that such a Fermi liquid structure can be
identified with Walecka's mean field model provided that BR scaling is suitably
implemented in the fluctuations.
In the work reported here, we will have to assume that we have
a nuclear matter  that comes out of an effective chiral action. Given
such a ground state containing no strange degrees of freedom, we would
like to study fluctuations along the strangeness direction and
determine if instability along that direction develops
signaling a phase transition.
We are therefore assuming that we can get the properties of normal nuclear
matter from phenomenology, that is, that nuclear matter is a
Fermi-liquid fixed point\cite{shankar,polchinski}. In principle,
a precise knowledge
of this ground state from a chiral effective Lagrangian at a nonperturbative
QCD level would
allow us to determine the coefficients that appear in the effective Lagrangian
with which to describe fluctuations
around the soliton background -- i.e.,
the Fermi liquid --and with which we could then compute all nuclear response
functions. At present such a derivation does not exist. In a recent paper
by Brown and one of the authors (BR91)\cite{brscaling},
it is assumed that in medium at a matter density $\rho\sim \rho_0$,
the {\it nuclear} effective field theory can be written in terms of
the medium-dependent coupling constants $g^\star$ and masses
of hadrons $m^\star$ while preserving the free-space structure of
a sigma model. This leads to the so-called Brown-Rho scaling. In
BR91\cite{brscaling}, the nonlinear sigma model implemented with trace anomaly
of QCD is used to arrive at the scaling law. The precise way that this
scaling makes sense is elaborated by Adami and Brown\cite{adamibrown} and
in the review (BR94)\cite{newbr94}. There have been numerous papers written
with some of the essential points of this scaling misinterpreted.

Given such an effective field theory, we can make a general argument on the
stability in various flavor directions of nuclear matter at high density.
This can be done along the line of arguments developed for condensed matter
physics by
Shankar\cite{shankar} and Polchinski\cite{polchinski} using renormalization
group flow. We sketch the essential argument following Lee, Rho and
Sin\cite{LRS}.

What we are interested in is whether the system in question develops
instability along the direction of strangeness and if so, by which
physical mechanism. This analysis will not give us the critical density.
The critical density will be calculated by using chiral perturbation
theory. For this purpose we will focus on the kaon frequency near
the electron chemical potential. By Baym's theorem\cite{baym},
one can identify
the kaon chemical potential associated with charge conservation,
$\mu_K$, with the electron chemical potential, $\mu_e$, which we shall simply
write $\mu$ in what follows. This means that we will be looking
at the vicinity of $\omega\sim \mu$ in the kaon dispersion formula.
We shall assume that
\be
|\omega-\mu| \ll \mu.
\ee
As mentioned, we assume
that nucleons in nuclear matter are in Fermi-liquid state with the Fermi
energy $\mu_F$ and the Fermi momentum $k_F$.  Define $\psi$ as the nucleon
field fluctuating around the Fermi surface
such that the momentum integral has a cut-off $\Lambda_N$,
\be
k_F -\Lambda_N < |\vec{k}| < k_F + \Lambda_N.
\ee
Kaons can interact with the nucleons through three-point functions of
the $KNN$ type (Yukawa interaction) and through four-point interactions
of the $KKNN$ type. We shall consider S-wave kaon-nucleon interactions,
for which the Yukawa interaction can be ignored. A generic action involving
the nucleon field $\psi$ and the kaon field
$\Phi$ can then be written, schematically, as
\be
S &=&  \int d\omega  d^3q \Phi^{*}(\omega, \vec{q})\left(\omega -
q^2/2\mu_K\right)\Phi(\omega, \vec{q})-\int d\omega  d^3q\, \tilde{M}_K
\Phi^*\Phi \nonumber\\
&& + \int (d\omega d^3q)^2 (d\epsilon d^3k)^2 h \Phi^*\Phi \psi^{\dagger}
 \psi \delta^4(\omega,\epsilon, \vec{q}, \vec{k})\nonumber\\
&& + \int d\epsilon d^3k \psi^{\dagger}\left(\epsilon -\epsilon(k))\right)\psi
    +g \int (d\epsilon d^3k)^4 \psi^{\dagger}\psi^{\dagger}\psi\psi
    \delta^4(\epsilon, \vec{k})
\label{toy1}
\ee
where $\tilde{M}=(M_K^2-\mu^2)/2\mu$ and $h$ and $g$ are constants. The
four-Fermi interaction with the coefficient $g$ stands for Fermi-liquid
interactions in nuclear matter. (In nuclear matter, one can have four such
terms because of the nucleon spin and isospin degrees of freedom. We need
not specify
them for our purpose.) This is a toy action but it is generic in that the
results of \chpt \ we will obtain below can be put into this form.

The renormalization group flow of this action can be analyzed in the
following way. Since we are assuming that nuclear matter is a Fermi-liquid
fixed point, fluctuations in the non-strange direction in the nucleon sector
are stable: The four-Fermi interaction $g$ is irrelevant or at best marginal.
Fluctuations in the strange direction involve the kaon field $\Phi$.
Suppose we have integrated out all the high-frequency modes above the
cut-off $\Lambda$ measured with respect to $\mu$. We are interested in the
stability of the system
under the renormalization group transformation $\Lambda\rightarrow s\Lambda$
($s<1$) as $s\rightarrow 0$. A scaling analysis shows that the
interaction term $h$ is
irrelevant while the ``mass term" $\tilde{M}$ is relevant. The renormalization
group-flow of the ``mass term" and the interaction term $h$ can be
readily written down and solved\cite{LRS} (with $t=-\ln\ s$),
\be
\tilde{M}(t)=(\tilde{M}_0-{Dh_0\over 1+a})e^{t} + {Dh_0\over 1+a}e^{-at}
\label{sol}
\ee
with
\be
h(t)=h_0 e^{-at},\ \ \ h_0\geq 0
\ee
where $D=\frac{3(1+\alpha^2)\alpha}{2\mu}\rho_N >0$, $\alpha= \Lambda/k_F >0$
and $a=1/2$. We see from Eq.(\ref{sol}) that as $s\rightarrow 0$ for which
$h\rightarrow 0$,  $\tilde{M}$ changes sign for some
$(\tilde{M}_0, h_0\geq 0)$. Thus although irrelevant, an attractive interaction
$h_0$ determines the direction of the mass flow whereas it is the ``mass term"
that drives the system to instability.

\subsection{\it Chiral Counting}
Armed with the general information on the instability in the strangeness
direction, we are now able to
calculate the critical density in \chpt.\ As mentioned,
we are to look at the instability in the kaon direction, so it suffices
for us to look at fluctuations around the Fermi-liquid state.
For this we need an effective chiral Lagrangian involving baryons as well
as Goldstone bosons. When baryons are present, \chpt \ is not as firmly
formulated as when they are absent\cite{leutwyler}. The reason is that
the baryon mass $m_B$ is $\sim \Lambda_\chi\sim 1$ GeV, the chiral
symmetry breaking scale. It is more expedient, therefore, to redefine
the baryon field so as to remove the mass from the baryon propagator
\be
B_v=e^{im_B\gamma\cdot v\; v\cdot x} P_+ B
\ee
where $P_+=(1+\gamma\cdot v)/2$ and write the baryon four-momentum
\be
p_\mu=m_B v_\mu +k_\mu
\ee
where $k_\mu$ is the small residual momentum indicating the baryon
being slightly off-shell.
When acted on by a derivative, the baryon field $B_v$ yields
a term of $O(k)$.
%------------------------------------------------
 The (octet) baryon propagator simplifies in heavy-baryon formalism to
$ i/v\cdot k $ that involves no gamma matrices. This simplifies
the loop calculation.
The spin operator $S_v^\mu$ is defined by,
$v\cdot S_v = 0$, $S_v^2 B_v =  -\frac 34 B_v$,
$ \{ S_v^\mu, S_v^\nu \} = \frac 12 (v^\mu v^\nu -g^{\mu\nu})$, and
$\left[ S_v^\mu, S_v^\nu \right] =
i\epsilon^{\mu\nu\alpha\beta} v_\alpha (S_v)_\beta$.
In the baryon rest frame, the spin operator $S_v$ reduces to the usual
spin operator $\vec\sigma/2$.

Chiral perturbation theory in terms of $B_v$ and
Goldstone bosons $(\pi\cdot\lambda/2)$ is known as
``heavy-baryon (HB) \chpt"\cite{HFF}. HB\chpt \ consists of making
chiral expansion in derivatives on Goldstone boson fields,
$\del_M/\Lambda_\chi$,
and on baryon fields, $\del_B/m_B$,  and in the quark mass matrix,
$\kappa {{\cal M}}/\Lambda_\chi^2$. In the meson sector, this is just what
Gasser and Leutwyler did for $\pi\pi$ scattering. In the baryon sector,
consistency with this expansion requires that the chiral counting be made
with $B^\dagger (\cdots)B$, not with $\bar{B} (\cdots)B$. This means that
in medium, it is always the baryon density $\rho (r)$ that comes in and
{\it not} the scalar density $\rho_s (r)$. This point seems to be
misunderstood by some workers in the field.

Following Weinberg\cite{weinberg}, we organize the chiral expansion
in power $Q^\nu$ where $Q$ is the characteristic energy/momentum scale
we are looking at ($Q<< \Lambda_\chi$) and
\be
\nu=4-N_n-2C+2L +\sum_i \Delta_i
\ee
with the sum over $i$ running over the vertices that appear in the graph
and
\be
\Delta_i=d_i +\frac 12 n_i -2.
\ee
Here $\nu$ gives the power of small momentum (or energy) for a process
involving $N_n$ nucleon lines, $L$ number of loops,
$d_i$ number of derivatives (or powers of meson mass) in the $i$th
vertex, $n_i$ number of nucleon lines entering into $i$th vertex and
$C$ is the number of separate connected pieces of the Feynman graph.
Chiral invariance requires that $\Delta_i\geq 0$, so that the leading
power is given by $L=0$, $\nu=4-N_N-2C$.

As an example, consider $KN$ scattering.
The leading term here is the tree graph with $\nu=1$ and with $N_n=C=1$.
The next order terms are $\nu=2$ tree graphs with $\Delta=1$ that involves
two derivatives or one factor of the mass matrix ${\cal M}$.
{}From $\nu=3$ on, we have loop graphs contributing
together with appropriate counter terms.

In considering kaon-nuclear interactions as in the case of kaon condensation,
we need to consider the case with $N_n\geq 2$ and $C\geq 2$. In dealing
with many-body system, one can simply fix $4-N_n$ and consider $C$ explicitly.
For instance if one has two nucleons (for reasons mentioned below, this
is sufficient, with multinucleon interactions being suppressed), then
we have $4-N_n=2$ but $C$ can be 2 or 1, the former describing a kaon
scattering on a single nucleon with a spectator nucleon propagating
without interactions and the latter a kaon scattering irreducibly on a
two-nucleon complex.  Thus intrinsic $n$-nucleon processes are suppressed
compared with $(n-1)$-nucleon processes by at least $O(Q^2)$.
This observation will be used later for arguing that four-Fermi interactions
are negligible in kaon condensation.  This is somewhat like the suppression
of three-body nuclear forces\cite{weinberg} and of three-body exchange
currents\cite{PMR} in chiral Lagrangians.

\subsection{\it Kaon-Nucleon Scattering}

Given a chiral Lagrangian, we need to first determine the parameters of the
Lagrangian from available phenomenology.
This is inevitable in effective field theories. We shall
first look at kaon-nucleon scattering at low energies. This was done
by Lee {\etal}\cite{LJMR,LBMR} which we summarize here. We shall compute
the scattering amplitude to one-loop order and this entails a Lagrangian
written to $O(Q^3)$ as one can see from the Weinberg counting rule.
Instead of writing it out in its full glory, we write it in a schematic
form as
\be
\L=\sum_i \L_i [B_v, U, \M]\label{lag}
\ee
where the subscript $i$ stands for $\nu$ relevant to the $KN$ channel.
Here $B_v$ stands for both octet and decuplet baryons and $U$ the Sugawara
form for octet Goldstone bosons.  For $KN$ scattering in free-space,
the Lagrangian is bilinear in the baryon field.
Details are given in Lee {\etal} \cite{LJMR,LBMR}.
Let us specify a few terms in (\ref{lag}) so as to streamline our discussion.
Focusing on S-wave scattering, $\L_1$ contains the leading order term that may
be described by the exchange of an $\omega$ between kaon and nucleon,
attractive for $K^-N$ and repulsive for $K^+ N$ and an isovector term
corresponding to the exchange of a $\rho$ meson. These terms are proportional
to the kaon frequency $\omega$. To next order,
$\L_2$ contains the ``$KN$ sigma term" proportional to $\Sigma_{KN}/f^2$
where $f$ is the pion decay constant and a term proportional to $\omega^2$
which may be saturated by decuplet intermediate states. The $\nu=3$ pieces are
counter terms that contain terms that remove divergences in the loop
calculations and finite terms that are to be determined from experiments.
The complete S-wave scattering amplitudes calculated
to the NNL order come out to be
\be
    a_0^{K^\pm p} &=& \frac{m_B}{4\pi f^2 (m_B+M_K)}
    \left[
    \mp M_K
    + (\bar d_s+\bar d_v) M_K^2 +\left\{ (L_s+L_v) \pm (\bar g_s +\bar g_v)
    \right\} M_K^3 \right]
  \nonumber\\
&&+\delta a_{\Lambda^\star}^{K^\pm p}\nonumber\\
  a_0^{K^\pm n} &=& \frac{m_B}{4\pi f^2 (m_B+M_K)}
    \left[ \mp \frac 12 M_K
    + (\bar d_s-\bar d_v) M_K^2 +\left\{ (L_s-L_v) \pm (\bar g_s -\bar g_v)
    \right\} M_K^3 \right]\nonumber\\
  \label{scattamp}\ee
  where $M_K$ is the kaon mass, $m_B$ the baryon (nucleon) mass,
$\bar d_s$ is the t-channel isoscalar contribution [${\cal O}(Q^2)$], and
$\bar d_v$ is the t-channel isovector one [${\cal O}(Q^2)$], both coming
from $\L_2$,  $L_s$($L_v$) is the finite crossing-even t-channel isoscalar
(isovector) finite one-loop contribution [$O(Q^3)$]
having the numerical values
  \be
    L_s M_K \approx -0.109 \; \fm, \ \ \ \
   L_v M_K    \approx +0.021 \; \fm
   \ee
and the quantity $\bar g_s (\bar g_v)$ is the crossing-odd t-channel isoscalar
(isovector) contribution [$O(Q^3)$] from one-loop plus counter terms in
$\L_3$.

To understand the role of the $\Lambda^\star$, we observe that the measured
scattering lengths are repulsive in all channels except
$K^- n$ \cite{BS}.
  \be
  a_0^{K^+p} = -0.31 \fm,& \;\;\;\;& a_0^{K^-p} = -0.67 +i 0.63 \fm
  \nonumber\\
  a_0^{K^+n} = -0.20 \fm,& \;\;\;\;& a_0^{K^-n} = +0.37 +i 0.57 \fm .
\label{expscatt}
  \ee
Although the experimental $K^- N$ scattering
lengths are given with error bars, the available $K^+ N$ data are not
very well determined. Since both are used in fitting the parameters of the
Lagrangian, we do not quote the error bars here and shall not
use them for fine-tuning. For our purpose,
we do not need great precision in the data as the results are
extremely robust against changes in the parameters.
The repulsion in $K^-p$ scattering cannot be explained from Eq.(\ref{scattamp})
without the $\Lambda^\star$ contribution.
In fact it is well known that the contribution of the
$\Lambda(1405)$ bound state
gives the repulsion required to fit empirical data for S-wave $K^-p$
scattering \cite{LJMR,LBM,LBMR,lambda1405}. As mentioned, we may introduce the
$\Lambda^\star$ as an elementary field. It takes the form
\be
   \delta a_{\Lambda^\star}^{K^\pm p} &=& - \frac{m_B}{4\pi f^2 (m_B+M_K)}
   \left[ \frac{ g_{\Lambda^\star}^2 M_K^2}{m_B \mp M_K -
m_{\Lambda^\star}}\right]
   \label{Lama}\ee
which is completely determined given experimental data on the coupling
  $ g_{\Lambda^\star}$ and the complex mass $m_{\Lambda^\star}$.

There are four unknowns $\bar{d}_{s,v}$, $\bar{g}_{s,v}$ in (\ref{scattamp})
which can be determined from four experimental (real part of)
scattering lengths Eq.(\ref{expscatt}).
The results are
   \be
   \bar d_s \approx 0.201  \fm, \;\; && \bar d_v \approx 0.013 \fm,
   \nonumber\\
   \bar g_s M_K \approx 0.008 \fm, \;\; && \bar g_v M_K \approx 0.002 \fm.
   \label{num}\ee

So far, no prediction is made. However given the parameters so fixed,
one can then go ahead and calculate the S-wave amplitude that enters
in kaon condensation. This amounts to going off-shell in the $\omega$
variable, that is, in the kinematics where $\omega\neq M_K$. In doing this,
one encounters an ambiguity due to the $\omega$ dependence of the
coefficients $\bar{d}$ which consist of the ``$KN$ sigma term" and ``$\omega^2$
term" which get compounded on-shell into one term. In the calculation
reported in Lee {\etal}\cite{LJMR,LBMR}, we chose to fix the ``$\omega^2$
term" by resonance saturation and leave the ``sigma term" to be fixed
by the on-shell data. The predicted off-shell amplitudes\cite{LJMR,LBMR}
agree reasonably with phenomenologically constructed off-shell amplitudes.
All the constants of the chiral Lagrangian bilinear in the baryon field
are thereby determined to $O(Q^3)$.

\subsection{\it Four-Fermi Interactions}

In medium, the chiral Lagrangian can have multi-Fermi interactions
as a result of ``mode elimination." Here we consider four-Fermi interactions,
ignoring higher-body interactions. We shall see that this is justified.

As stated above, we need to focus on four-Fermi interactions that involve
strangeness degrees of freedom. Nonstrange four-Fermi interactions are subsumed
in the Fermi-liquid structure of normal nuclear matter. For S-wave kaon-nuclear
interactions, we only have the $\Lambda (1405)$ to account for. There are
only two terms,
 \be
    {\cal L}_{4-fermion} &=&
   C_{\Lambda^\star}^S \bar{\Lambda}^\star_v \Lambda^\star_v
  \Tr \bar B_v B_v +  C_{\Lambda^\star}^T
  \bar{\Lambda}^\star_v \sigma^k \Lambda^\star_v
     \Tr \bar B_v \sigma^k B_v\label{fourfermi}
    \ee
where $C_{\Lambda^\star}^{S,T}$ are the
dimension $-2$ ($M^{-2}$) parameters to be fixed
empirically and $\sigma^k$ acts on baryon spinor. We shall now describe
how to fix these two parameters from kaonic atom data.

\begin{figure}
 \centerline{\epsfig{file=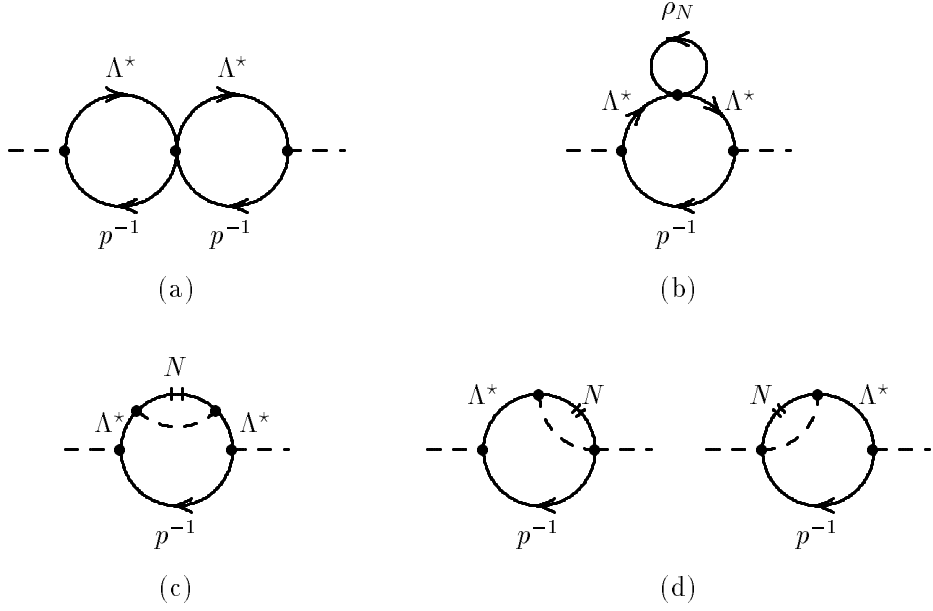,height=8cm}}
\caption[selfenergy] {The in-medium two-loop kaon self-energy
involving $\Lambda (1405)$. Figures a and b contain the constants
of the four-Fermi interaction and figures c and d are Pauli corrections}
\label{selfenergy}
\end{figure}

In order to confront kaonic atom data, we need to calculate the kaon
self-energy $\Pi$ in nuclei. The off-shell amplitude determined above
gives the so-called ``impulse" term
 \be
    \Pi^{imp}_K(\omega) &=& -\left( \rho_p  {\cal T}^{K^-p}_{free}(\omega)
        +\rho_n  {\cal T}^{K^-n}_{free}(\omega) \right)\label{self1}
    \ee
where ${\cal T}^{KN}$ is the off-shell S-wave KN transition matrix.
(The amplitude ${\cal T}^{KN}$ taken on-shell, {\it i.e.},
$\omega=M_K$, and the scattering length
$a^{KN}$ are related by $ a^{KN} = \frac{1}{4\pi (1+M_K/m_B)}
{\cal T}^{KN} $.)  Medium corrections to this ``impulse" term,
obtained from one-loop graphs by replacing the free-space
nucleon propagator by the in-medium propagator, shall be denoted as
\be
 - \left(\rho_p  \delta {\cal T}^{K^-p}_{\rho_N}(\omega)
        +\rho_n \delta {\cal T}^{K^-n}_{\rho_N}(\omega) \right).\label{deself}
\ee
These two terms (\ref{self1}) and (\ref{deself})
are completely determined by the parameters fixed above.
The new parameters of the four-Fermi interaction come into play in the
first two self-energy graphs of Fig.\ref{selfenergy}
(the last two graphs do not
involve four-Fermi interactions but enter at the same order; they are
free of unknown parameters),
 \be
    \Pi_{\Lambda^\star}(\omega) &=& - \frac{g_{\Lambda^\star}^2}{f^2}
        \left(\frac{\omega}{\omega+m_B-m_{\Lambda^\star}} \right)^2
    \left\{ C_{\Lambda^\star}^S \rho_p \left( \rho_n +\frac 12 \rho_p \right)
    -\frac 32 C_{\Lambda^\star}^T \rho_p^2 \right\}
    \nonumber\\
    && + \frac{g_{\Lambda^\star}^2}{f^4} \rho_p
    \left(\frac{\omega}{\omega+m_B-m_{\Lambda^\star}} \right)
    \omega^2 \left\{  \left( 2 \Sigma_K^p (\omega) +\Sigma_K^n (\omega) \right)
    \vphantom{\frac 12} \right.
    \nonumber\\
    && \;\;\;\;\;\;\;\;\;\;\;\;\;\;\; \;\;\;\;\;\;\; \left.
     -g_{\Lambda^\star}^2
    \left(\frac{\omega}{\omega+m_B-m_{\Lambda^\star}} \right)
     \left( \Sigma_K^p (\omega) +\Sigma_K^n (\omega) \right)
    \right\}
    \label{pilambda}\ee
where ${g}_{\Lambda^\star}$ is the renormalized $KN\Lambda^\star$ coupling
constant determined in Lee {\etal}\cite{LJMR,LBMR}
and $\Sigma_K^N (\omega)$ is a known integral that depends on proton and
neutron densities and $M_K$.
Note that while the second term of (\ref{pilambda})
gives repulsion corresponding to a Pauli quenching,
the first term can give either attraction or repulsion depending on
the sign of $(C_{\Lambda^\star}^S [\rho_n+\frac 12 \rho_p]-
\frac 32 C_{\Lambda^\star}^T\rho_p)$. For symmetric nuclear matter, only the
combination  $(C_{\Lambda^\star}^S - C_{\Lambda^\star}^T)$ enters in the
self-energy. This is an important element for kaonic atom.

The complete self-energy to in-medium two-loop order
is then
    \be
    \Pi_K(\omega) &=& -\left( \rho_p  {\cal T}^{K^-p}_{free}(\omega)
        +\rho_n  {\cal T}^{K^-n}_{free}(\omega) \right)
        - \left(\rho_p  \delta {\cal T}^{K^-p}_{\rho_N}(\omega)
        +\rho_n \delta {\cal T}^{K^-n}_{\rho_N}(\omega) \right)\nonumber\\
       && +\Pi_{\Lambda^\star}(\omega)\label{self2}.
    \ee

We now turn to fixing the constants of the four-Fermi interactions
based on the recent analysis of kaonic atoms by Friedman, Gal
and Batty \cite{kaonicatom}. For later purpose we shall parameterize
the proton and neutron densities by the proton fraction $x$
and the nucleon density $u=\rho/\rho_0$ as
    \be
    \rho_p = x\rho\; ,\;\; \rho_n =(1-x) \rho\; ,\;\; \rho = u\rho_0.
    \ee
Now Friedman {\etal}\cite{kaonicatom} found from their analysis that the
optical potential for the $K^-$ in medium has an attraction of the order of
    \be
    \Delta V\equiv M_K^\star-M_K
    \approx -(200\pm 20)\ \ {\rm  MeV}\ \ at \;\; u=0.97
    \ee
    with
    \be
M_K^\star \equiv\sqrt{M_K^2+\Pi_K}.
\ee
This implies approximately for $x=1/2$
    \be
     ( C_{\Lambda^\star}^S -C_{\Lambda^\star}^T) f^2\approx 20.
\label{cvalue}
    \ee
Friedman {\etal} \cite{kaonicatom} note that their ``nominal" optical
potential gives an attraction of order of 800 MeV when extrapolated
to three times the normal density. We show in Table \ref{Table1}
 what our theory predicts
at higher densities than normal.\footnote{
The numerical values in Tables~\ref{Table1} and \ref{Table2}
are slightly modified from the previous results in \cite{LRp} which had
numerically small errors. Specifically, the
modifications in $\delta {\cal T}^{K^-N}_{\rho_N}$, Eq.~(G.3) of \cite{LBMR},
$$
    D_{6,ij}^N \longrightarrow  - D_{6,ij}^N, \;\;\;\;  M_i^2 \Sigma_i^{N}(0)
\longrightarrow
      \frac{1}{2\pi^2} \int_0^{k_{F_N}} d|\vec k|
      \frac{|\vec k|^4}{M_i^2+|\vec k|^2},
$$
are responsible for the slight changes in the numerical values.
We have verified that our new results (See Eq.~(G.1) of \cite{LBMR})
 satisfy the chiral symmetry constraint of Mei{\ss}ner et al.\cite{MOP}.
}
At $u=3$, the net attraction is only about 1.7 times the one at $u=1$.

\begin{table}
    $$
    \begin{array}{|c||c|c|}
    \hline
     u & M_K^* & \Delta V  \\
    \hline
    \hline
 0.5& 354.9& -140.1 \\ \hline
 1.0& 294.4& -200.6 \\ \hline
 1.5& 249.2& -245.8 \\ \hline
 2.0& 211.7& -283.3 \\ \hline
 2.5& 180.5& -314.5 \\ \hline
 3.0& 153.9& -341.1 \\ \hline
 3.5& 130.8& -364.2 \\ \hline
 4.0& 113.2& -381.8 \\ \hline
    \end{array}
    $$
\caption{$K^-$ effective mass($M_K^*$) and the attraction
    ($\Delta V\equiv M_K^\star -M_K$ )
    in symmetric nuclear matter ($x=0.5$) as function
    of density $u$ in unit of MeV for
     $(C_{\Lambda^\star}^S-C_{\Lambda^\star}^T) f^2 =20$.  }
\label{Table1}
\end{table}

Equation (\ref{pilambda}) shows that for symmetric nuclear matter ($x=1/2$),
the combination $(C_{\Lambda^\star}^S +C_{\Lambda^\star}^T)$ does not enter
into the
self-energy formula. In order to extract it as needed for non-symmetric
system as in compact star matter, we need information for nuclei with
$x\neq 1/2$. This can be done from the results of Friedman {\etal}
by noting that our self-energy is nonlinear in $x$, so
\be
\frac{\del\Delta V}{\del x} (C_{\Lambda^\star}^S,\rho\approx \rho_0)|_{x=1/2}
\approx 400 \;  b_1/b_0 \ \ {\mbox MeV}
\ee
where $b_{0,1}$ are the constants given by Friedman {\etal}.  This relation
determines the coefficient $C_{\Lambda^\star}^S$. The result is shown in
Table \ref{Table2} (first three columns).

Friedman {\etal}\cite{kaonicatom} find the acceptable value to be
$b_1/b_0= -0.56\pm 0.82$. But there is one point which needs to be discussed
in interpreting this number in the context of our theory.
The constant $C_{\Lambda^\star}^S$ shifts linearly the effective
in-medium mass of $\Lambda (1405)$, with the mass shift being given by
\be
\delta m_{\Lambda^\star}=\sum_{i=a,b}\delta
 \Sigma^{(i)}_{\Lambda^\star} (\omega=m_{\Lambda^\star}-m_B)
\ee
where
\be
\delta\Sigma_{\Lambda^\star}^{(a)} (\omega)
&=&-\frac{g_{\Lambda^\star}^2}{f^2}\omega^2\left(
\Sigma^p_K (\omega) +\Sigma^n_K (\omega)\right)\nonumber\\
\delta\Sigma_{\Lambda^\star}^{(b)} (\omega) &=&
-C_{\Lambda^\star}^S (\rho_p+\rho_n).
\ee
For nuclear matter density $u=1$ and $x=1/2$, the shift is
\be
\delta m_{\Lambda^\star} (u,x,y)\approx [62-150.3\times y] \ \ {\rm MeV}
\label{lambdamass}
\ee
with $y=C^S_{\Lambda^\star} f^2$. It seems highly unlikely that
the $\Lambda (1405)$ will be shifted by hundreds of MeV in nuclear matter.
This means that
$y$ must be of $O(1)$, and {\it not} $O(10)$.  For $y=0.41$
which corresponds to $b_1/b_0\approx -0.4$, there is no shift at normal matter
density.  We believe this is a reasonable value. In fact,
$y=0$ is also acceptable. It would be interesting to
measure the shift of $\Lambda (1405)$ to fix the constant
$C_{\Lambda^\star}^S$ more precisely although its precise magnitude
seems to matter only a little for kaonic atoms and as it turns out, negligibly
for kaon condensation.

Let us comment briefly on the role of multi-Fermion Lagrangians.
The Weinberg counting rule shows that the four-Fermi interactions
are suppressed by $O(Q^2)$ relative to the terms involving bilinears of
Fermi fields. In general $n$-Fermi interactions will be suppressed
by the same order relative to $(n-1)$-Fermi interactions. In considering
kaon condensation, what this means in conjunction with the
renormalization-group flow argument, is that $n$-Fermi interactions
with $n\geq 4$ are irrelevant in the RGE sense, and hence unimportant
for condensation. The situation with the kaonic atom data is a bit different.
While the strength of the four-Fermi interaction, $y$, is not important
(this can be seen in Lee {\etal}\cite{LBMR}, Table 3), its presence is
essential for the attraction that seems to be required. This is in contrast
to the kaon condensation which is driven by the ``mass flow" with four-Fermi
interactions being irrelevant in the RGE sense.

\begin{table}
$$
\begin{array}{|c|r|r|c|c|c|}
\hline
 & & & \multicolumn{3}{c|}{ u_c } \\ \cline{4-6}
y=C_{\Lambda^\star}^S f^2 & \partial\Delta V/\partial x & b_1/b_0
& F(u)=\frac{2u^2}{1+u} & F(u)= u &  F(u)=\sqrt u \\
\hline
50  &  125.44 \MeV &  0.314 &      2.25 &      2.50 &      2.97 \\
40  &  64.77 \MeV &  0.162 &      2.33 &      2.58 &      3.08 \\
30  &  4.10 \MeV &  0.010 &      2.42 &      2.69 &      3.22 \\
20  &  -56.58 \MeV & -0.141 &      2.54 &      2.84 &      3.41 \\
10  & -117.35 \MeV & -0.293 &      2.71 &      3.05 &      3.71 \\
\hline \hline
0.41 & -175.43 \MeV & -0.439 &  2.98  & 3.43 &  4.28 \\
\hline \hline
0   & -177.92 \MeV & -0.445 &      2.99 &      3.45 &      4.32 \\
-10 & -238.59 \MeV & -0.596 &      3.60 &      4.85 & \sim 6.41 \\
\hline
\end{array}
$$
\caption[tabel2]{Determination of ${C_\Lambda^\star}^S$ from the kaonic atom
data\protect{\cite{kaonicatom}} and the critical density (obtained with the
constant so determined) for kaon condensation
for various forms of symmetry energy $F(u)$ and
$({C_\Lambda^\star}^S-{C_\Lambda^\star}^T) f^2=20$.
$y=0.41$ corresponds to no $\Lambda (1405)$ mass shift in medium
at the normal matter
density.}
\label{Table2}
\end{table}

\section{Kaon Condensation}
\subsection{\it Equation of State and Critical Densities}
We have now all the ingredients needed to calculate the critical density
for negatively charged kaon condensation in dense nuclear star matter.
For this, we will follow the procedure given in work of Thorsson, Prakash
and Lattimer (TPL)\cite{TPL}.
As argued by Brown, Kubodera and Rho\cite{BKR}, we need not consider pions
when electrons with high chemical potential can trigger condensation through
the process $e^-\rightarrow K^- \nu_e$. Thus we can focus on the spatially
uniform condensate
    \be
    \langle K^-\rangle =v_K e^{-i\mu t}.
    \ee
The energy density $\tilde\epsilon$ -- which is related to the
effective potential in the standard way -- is given by,
    \be
    \tilde \epsilon (u,x,\mu, v_K) &=& \frac 35 E_F^{(0)} u^{\frac 53} \rho_0
        +V(u) +u\rho_0 (1-2x)^2 S(u) \nonumber\\
    &&-[\mu^2 -M_K^2 -\Pi_K (\mu,u,x)]
        v_K^2+ \sum_{n\ge 2} a_n(\mu,u,x) v_K^n \nonumber\\
    && +\mu u\rho_0 x +\tilde\epsilon_e +\theta(|\mu|-m_\mu)\tilde \epsilon_\mu
    \label{effen}\ee
where $E_F^{(0)}=\left( p_F^{(0)}\right)^2/2m_B$ and
$p_F^{(0)}=(3\pi^2\rho_0 /2)^{\frac 13}$ are, respectively,
 Fermi energy and momentum at nuclear density. The
$V(u)$ is a potential for symmetric nuclear matter
as described by Prakash {\etal}\cite{PAL}
which is presumably subsumed in contact four-Fermi
interactions (and one-pion-exchange -- nonlocal -- interaction)
in the non-strange sector as mentioned above. It will affect
the equation of
state in the condensed phase but not the critical density, so we will
drop it from now on. The nuclear symmetry energy $S(u)$ -- also
subsumed in four-Fermi interactions in the non-strange sector -- does
play a role as we know from Prakash {\etal}\cite{PAL}:
Protons enter to neutralize the
charge of condensing $K^-$'s making the resulting compact star
``nuclear" rather than neutron star as one learns in standard astrophysics
textbooks. We take the form advocated by Prakash {\etal}\cite{PAL}
    \be
    S(u) &=& \left(2^{\frac 23}-1\right) \frac 35 E_F^{(0)}
        \left(u^{\frac 23} -F(u) \right) +S_0 F(u)
    \ee
where $F(u)$ is the potential contributions to the symmetry energy and
$S_0 \simeq 30 MeV$ is the bulk symmetry energy parameter.
We use three different forms of $F(u)$\cite{PAL}
    \be
    F(u)=u\;,\;\; F(u) =\frac{2u^2}{1+u}\;,\;\; F(u)=\sqrt u.
    \label{SE}
    \ee
The contributions of the filled Fermi seas of electrons and muons
are\cite{TPL}
    \be
    \tilde \epsilon_e &=& -\frac{\mu^4}{12\pi^2} \nonumber\\
    \tilde \epsilon_\mu &=& \epsilon_\mu -\mu \rho_\mu
    = \frac{m_\mu^4}{8\pi^2}\left((2t^2+1) t\sqrt{t^2+1}
    -\ln(t^2+\sqrt{t^2+1} )
    \right) -\mu \frac{p_{F_\mu}^3}{3\pi^2}
    \ee
where $p_{F_\mu} =\sqrt{\mu^2-m_\mu^2}$ is the Fermi momentum and $t=p_{F_\mu}
/m_\mu$.

\begin{table}
   $$
   \begin{array}{|c||c|c|c|c|}
   \hline
g_{\Lambda^\star}^2 &
   (C_{\Lambda^\star}^S -C_{\Lambda^\star}^T) f^2
   & C_{\Lambda^\star}^S f^2 =100
   &C_{\Lambda^\star}^S f^2 =10 &C_{\Lambda^\star}^S f^2 =0 \\
   \hline \hline
     &    1 &  2.25 &  3.29 &  4.91 \\ \cline{2-5}
0.25 &   10 &  2.25 &  3.16 &  3.76 \\ \cline{2-5}
     &  100 &  2.18 &  2.67 &  2.79 \\ \hline
  \end{array}
   $$
   \caption{Critical density $u_c$ in in-medium two-loop chiral
         perturbation theory for $F(u)=u$.}
\label{table7}
\end{table}

The ground-state energy prior to kaon condensation
is then obtained by extremizing the energy density $\tilde\epsilon$
with respect to  $x$, $\mu$ and $v_K$:
    \be
    \left. \frac{\partial\epsilon}{\partial x}\right |_{v_K=0}=0 \;,\;\;
    \left. \frac{\partial\epsilon}{\partial \mu}\right |_{v_K=0}=0 \;,\;\;
    \left. \frac{\partial\epsilon}{\partial v_K^2}\right |_{v_K=0}=0
    \ee
from which we obtain three equations corresponding, respectively,
to beta equilibrium, charge neutrality and dispersion relation.
The critical density so obtained is given for  three different $F(u)$'s
in Table \ref{Table2}, and for various ranges of parameters in
Table \ref{table7}. The result is
\be
2< u_c\lsim 4.\label{ineq}
\ee

We note that the largest sensitivity is associated with the part that is
not controlled by chiral symmetry, namely the density dependence of the
symmetry energy function $F(u)$. This uncertainty reflects the part of
interaction that is not directly given  by chiral Lagrangians, that is,
the part leading to normal nuclear matter. This is the major short-coming of
our calculation.

Related to this issue is BR scaling. As we argued, were we able to
derive nuclear matter from effective chiral Lagrangians, we would have
parameters of the theory determined at that point reflecting
the background around which fluctuations are to be made.
The BR scaling was proposed in that spirit but with a rather strong
assumption: That a sigma model governs dynamics in medium as in free space
with only coupling constants and masses scaled a function of density.
Up to date, no derivation of this scaling from basic principles
has been made. In this sense, we might consider it as a conjecture although
there is strong support for it from Walecka phenomenology in mean field
as discussed in \cite{newbr94}.
Suppose we apply
BR scaling. The only way the procedure can make sense is to apply the
scaling argument to the tree order terms, but not to the loop corrections.
The result of this procedure is significant in that the critical density
is brought down in an intuitively plausible way to about $u_c\sim 2$,
with very little dependence on parameters, loop
corrections and multi-Fermi interactions. Thus slightly modified from
(\ref{ineq}), we arrive at the announced result
\be
2\lsim  u_c\lsim 4.\label{ineqq}
\ee

\begin{figure}
\centerline{\epsfig{file=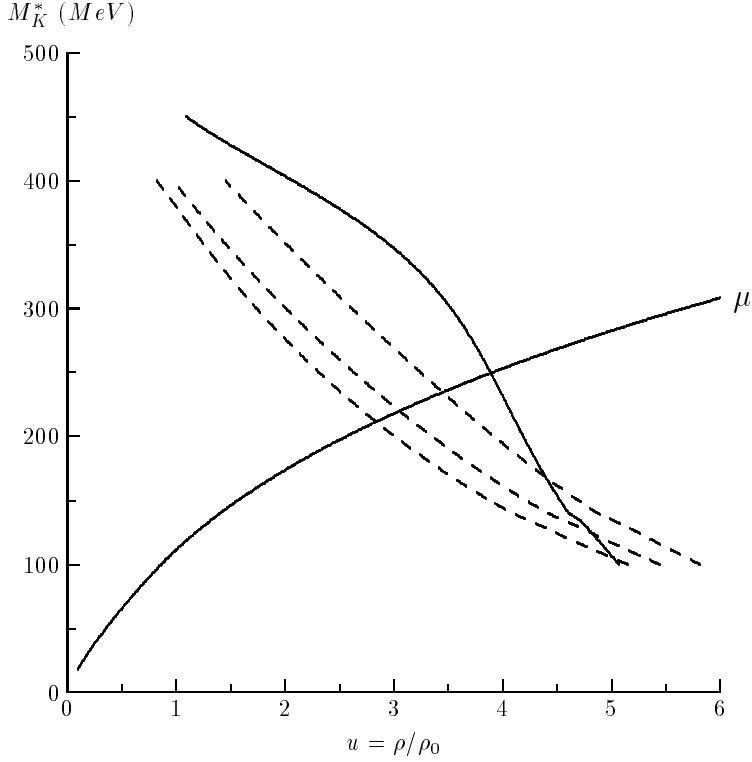,height=10cm}}
\caption{
    Plot of the quantity $M^\star_K$ obtained from
    the dispersion formula $D^{-1}(\mu,u) =0$ vs. the chemical
    potential $\mu$ prior to kaon condensation for $g_{\Lambda^\star}^2=0.25$
    and $F(u)=u$.
    The solid line corresponds to impulse approximation and
    the dashed lines to the in-medium two-loop results for
    $(C_{\Lambda^\star}^S-C_{\Lambda^\star}^T) f^2 =20$ and
    $C_{\Lambda^\star}^S  f^2 = 20, 10, 0$ respectively from the left.
    The point at which  the chemical potential $\mu$
    intersects $M_K^\star$ corresponds to the critical point.
}\label{fig9}
\end{figure}

\subsection{\it Irrelevance of $\Lambda^\star$ to Kaon Condensation}

\begin{figure}
\centerline{\epsfig{file=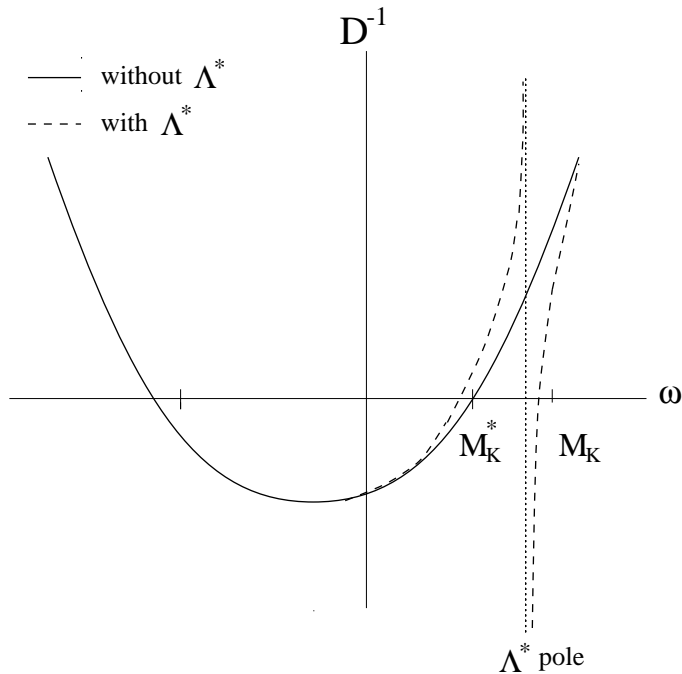,height=9cm}}
\caption{Plot of $D^{-1} =\mu^2 -M_K^2-\Pi(\mu)$}
\label{lambdairr}
\end{figure}

To see which modes are involved in S-wave kaon condensation, we consider
the dispersion formula at tree order,
\be
D^{-1}(\omega) =\omega^2 -M_K^2-\Pi(\omega).
\ee
As shown in Fig.~\ref{lambdairr} by the solid line, the kaon ``effective mass"
$M_K^*$ is reduced mainly by the KN sigma term when there are
no $\Lambda^\star$
contributions. If we turn on the $\Lambda^\star$ coupling, there
will be additional attractions. However since the effective mass $M_K^*$ lies
far from the $\Lambda^\star$-pole contribution, the resulting magnitude of
the attraction is small, i.e., the $M_K^*$ remains nearly unmodified.
Furthermore, since the $\Lambda^\star$-pole is far outside of $M_K^*$,
the condensed kaon mode remains the same independently of the $\Lambda^\star$.

Summarizing the results, $\Lambda^\star$ may be  crucial for understanding
the KN scattering and kaonic atom data, but is irrelevant to determining
the kaon condensation. The critical densities for wide ranges of
the $\Lambda^\star$ coupling in Table \ref{table7} confirm the unimportance
of $\Lambda^\star$ contribution.

%--------------------------------------------------
\section{Kaons on the Hypersphere}

\begin{figure}
\centerline{\epsfig{file=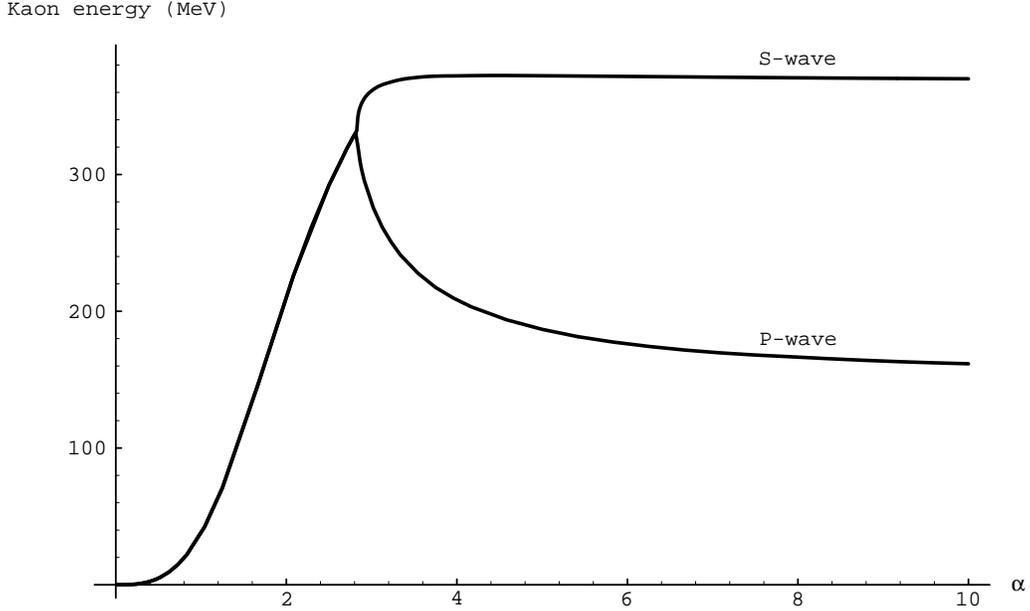,height=8cm}}
\caption[skyr]{Kaon energy (MeV) vs. $\alpha$\protect{\cite{Westerberg}}.}
\label{skyr}
\end{figure}

Recently, using the idea of Manton\cite{Manton} for simulating density
effects,
Westerberg\cite{Westerberg} explored S-wave kaon condensation
in the bound-state approach to the Skyrme model on a 3-sphere.
The spatial metric in a  hypersphere of radius $a$ is
\be
ds^2 = a^2 (d\rho^2 +\sin^2\rho (d\theta^2 +\sin^2\theta d\phi^2) )
\ee
where the possible ranges of three angular coordinates are
\be
0 \le\rho, \;\; \theta \le \pi,\;\; 0\le \phi \le 2\pi.
\ee
The baryon number density is given by the inverse volume of
the hypersphere
\be
\rho =\frac{1}{2\pi^2 a^3} \equiv \frac{e^3 F_\pi^3}{2 \pi^2\alpha^3}
\label{dens}\ee
where $\alpha=a e F_\pi$ with $F_\pi = 129 MeV$ and $e=5.45$.
The kaon energy is shown in Fig.~\ref{skyr} which also shows that
the chiral phase  transition occurs at $\alpha=\alpha_c=2\sqrt 2$.
For $\alpha > \alpha_c$,  the P-wave and S-wave kaons have different
masses, the difference in mass representing roughly the mass difference
between $\Lambda (1405)$ and $\Lambda (1116)$.
For $\alpha <\alpha_c$, the S-wave and P-wave
kaons become degenerate, with the kaon condensation occurring
in the regime $1<\alpha<2\sqrt 2$.

Solving the equation of state for the electron chemical potential,
Westerberg\cite{Westerberg} found the critical density to be at $\alpha=1.58$
corresponding to $\rho\approx 3.7 \rho_o$, Eq.(\ref{dens}). This
falls within our predicted range $(2\sim 4) \rho_0$. However an unsatisfactory
aspect of this result is that the kaon condensation sets in {\it after}
-- and {\it not} before as one expects -- the chiral phase transition.

\section{Discussion}

Introducing $\Lambda^\star$ up to order $Q^3$, we obtain the critical
density close to that of Kaplan and Nelson\cite{KN}, and confirm that the
KN sigma term, as in original approach of \cite{KN},
 is essential for kaon condensation. This result is further supported by
a renormalization-group flow argument as well as by
the recent bound-state approach
to the Skyrme model.

Given that kaon condensation occurs at a low enough density as predicted here,
the Bethe-Brown scenario seems very plausible.
However
whether or not the Bethe-Brown scenario\cite{BB} of compact star formation
is fully supported by the chiral Lagrangian approach will have to await the
calculation of the equation of state at in-medium two-loop order,
which is in progress. Our conjecture is that to the extent that our
work confirms the original Kaplan-Nelson calculation\cite{KN},
the compact star
properties calculated previously at the tree level\cite{TPL} would
come out qualitatively unmodified in the higher-order chiral perturbation
theory.

%--------------------------------------------------

\section*{Acknowledgments}
We would like to thank G.E. Brown, H.K. Lee, D.-P. Min and
S.-J. Sin for discussions. The work of CHL is supported in part by the
Korea Science and Engineering Foundation through the CTP of SNU and in
part by the Korea Ministry of Education under Grant No. BSRI-94-2418.
Part of this paper was written at the INT while MR was participating
in the INT95-1 program on ``Chiral symmetry in hadrons and nulcei."

\end{document}